\documentclass[a4paper]{jpconf}
\usepackage{graphicx}
\usepackage{times}
\usepackage{epsfig}
\usepackage{color}
\usepackage{comment}

\def\beq{\begin{equation}}
\def\eeq{\end{equation}}
\def\bea{\begin{eqnarray}}
\def\eea{\end{eqnarray}}

\def\nn{\nonumber}
\def\Eq#1{Eq.~(\ref{#1})}

\begin{document}
\title{Charge asymmetries of top quarks: a window to new physics at hadron colliders}

\author{Paola Ferrario$^1$ and Germ\'an Rodrigo$^2$}

\address{Instituto de F\'{\i}sica Corpuscular, 
CSIC-Universitat de Val\`encia,
Apartado de Correos 22085, 
E-46071 Valencia, Spain.}

\ead{$^1$ paolafer@ific.uv.es, $^2$ german.rodrigo@ific.uv.es}

\begin{abstract}

With the next start of LHC, a huge production of top quarks is expected. There are several models 
that predict the existence of heavy colored resonances 
decaying to top quarks in the TeV energy range. A peak in the differential cross 
section could reveal the existence of such a resonance, but this 
is experimentally challenging, because it requires selecting data samples 
where top and antitop quarks are highly boosted. Nonetheless, the 
production of such resonances might generate a sizable charge asymmetry of top versus 
antitop quarks. We consider a toy model with general flavour 
independent couplings of the resonance to quarks, of both vector and axial-vector kind.
The charge asymmetry turns out to be a more powerful observable to detect 
new physics than the differential cross section, because its highest 
statistical significance is achieved with data samples of top-antitop quark pairs 
of low invariant masses. 

\end{abstract}



\section{Introduction}

The CERN Large Hadron Collider (LHC) will start-up very soon
colliding protons to protons. The full $\sqrt{s}=14$~TeV design energy run will operate
with an initial low luminosity of ${\cal L}=10^{33}$cm$^{-2}$s$^{-1}$
(equivalent to $10$~fb$^{-1}$/year integrated luminosity) and then it will pass to a second phase of ${\cal L}=100$~fb$^{-1}$/year.
The production cross section of top-antitop quark pairs at LHC
is about $430$~pb at $10$~TeV, and $950$~pb 
at 14~TeV~\cite{Cacciari:2008zb}. The LHC will produce even
in the first phase of operation a sample of $t\bar t$-pairs 
equivalent to the sample already collected at Tevatron 
during its whole life, and millions of top-antitop quark pairs 
in the next runs.
This will allow not only to measure better some of the properties 
of the top quark, such as mass and cross section, but also 
to explore with unprecedented huge statistics the existence 
of new physics at the TeV energy scale in the top quark sector.

\begin{figure}[th]
\begin{center}
\includegraphics[width=9cm]{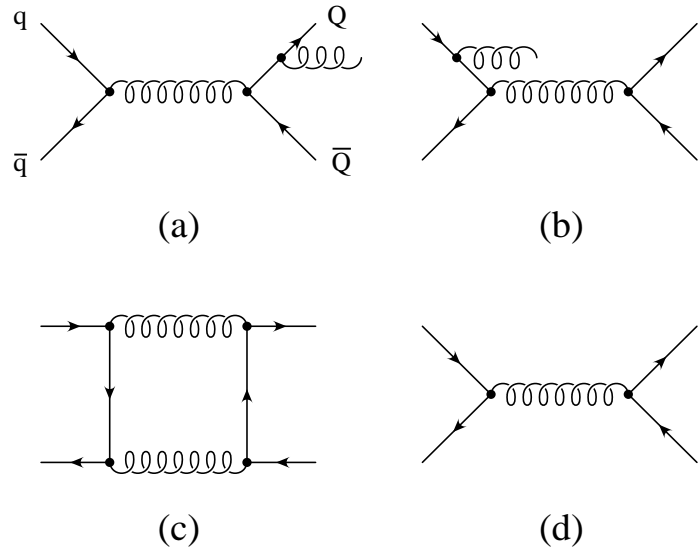}
\caption{Origin of the QCD charge asymmetry in hadroproduction
of heavy quarks: interference of final-state
(a) with initial-state (b) gluon bremsstrahlung,
plus interference of the double virtual gluon exchange (c)
with the Born diagram (d). Only representative diagrams are shown.}
\label{fig:qqbar}
\end{center}
\end{figure}

At leading order in the strong coupling $\alpha_s$ the differential 
distributions of top and antitop quarks are identical. 
This feature changes, however, due to higher order 
corrections~\cite{mynlo}, which predict at ${\cal O}(\alpha_s^3)$
a charge asymmetry of top versus antitop quarks.
The inclusive charge asymmetry receives contributions from two reactions:
radiative corrections to quark-antiquark annihilation (Fig.~\ref{fig:qqbar})
and interference between different amplitudes contributing
to gluon-quark scattering $gq \to t \bar{t}q$ and 
$g\bar{q} \to t \bar{t}\bar{q}$. The latter contribution is, 
in general, much smaller than the former. 
Gluon-gluon fusion remains charge symmetric.
At Tevatron, this asymmetry is equivalent to a forward--backward asymmetry and QCD predicts that the size of the inclusive charge asymmetry 
is 5 to 8\%~\cite{mynlo,Antunano:2007da,Bowen:2005ap}, 
with top quarks (antitop quarks)
more abundant in the direction of the incoming proton (antiproton). This is in agreement with the experimental data~\cite{newcdf,d0}, but more statistic is needed to reduce the errors.

At LHC, the total forward--backward asymmetry vanishes trivially
because the proton-proton initial state is symmetric. Nevertheless,
a charge asymmetry is still visible in suitably defined 
distributions~\cite{mynlo}. Top quark production at LHC is dominated by gluon-gluon 
fusion ($84~\%$ at $10$~TeV, and $90~\%$ at $14$~TeV), which
is charge symmetric under higher order corrections. 
The charge antisymmetric contributions to top quark 
production are thus screened at LHC 
due to the prevalence of gluon-gluon fusion. 
This is the main handicap for that measurement.
The amount of events initiated by gluon-gluon collisions can nevertheless
be suppressed with respect to the $q\bar q$ and $gq(\bar q)$ processes, 
the source of the charge asymmetry, by introducing a lower cut 
on the invariant mass of the top-antitop quark system $m_{t\bar t}$;
this eliminates the region of lower longitudinal momentum 
fraction of the colliding partons, 
where the gluon density is much larger than the quark densities. 
The charge asymmetry of the selected data samples is then enhanced,
although at the price of lowering the statistics. This is, in principle, 
not a problem at LHC, where the high luminosity will compensate 
by far this reduction. 

Several models predict the existence of heavy colored resonances 
decaying to top quarks that might be observed at the 
LHC~\cite{chiralcolor,Bagger:1987fz,Sehgal,Choudhury:2007ux,KK,Randall:1999ee,Dicus:2000hm,Agashe:2006hk,Lillie:2007yh,Djouadi:2007eg,Carone:2008rx,Frederix:2007gi}. 
Those resonances will appear as a peak in the invariant mass
distribution of the top-antitop quark pair located at the mass of the new 
resonance. 

Some of those exotic gauge bosons, such as the 
axigluons~\cite{chiralcolor,Bagger:1987fz}, 
might generate at tree-level a charge asymmetry too, through 
the interference with the $q\bar q \to t \bar t$ 
Standard Model (SM) amplitude~\cite{Antunano:2007da,Sehgal,Choudhury:2007ux}.
Gluon-gluon fusion to top quarks stays, at first order, 
unaltered by the presence of new interactions because a pair 
of gluons do not couple to a single extra resonance 
in this kind of models~\cite{Bagger:1987fz,Dicus:2000hm}. 

To discover those resonances, hence, it is necessary to select 
top-antitop quark events with large invariant masses; i.e. in 
the vicinity of the mass of the new resonance. 
A sizable charge asymmetry can also be obtained only if 
gluon-gluon fusion is sufficiently suppressed, that isº
at large values of $m_{t\bar t}$. 
Because the top quarks of those data samples will be produced
highly boosted, they will be observed as a single monojet. 
The standard reconstruction algorithms that are based on the reconstruction 
of the decays products, however, loose efficiency very rapidly at high 
transverse momentum. For $p_T > 400$~GeV new identification 
techniques are necessary. This has motivated many recent 
investigations~\cite{Kaplan:2008ie,Thaler:2008ju,Vos,Almeida:2008yp} aimed at 
distinguishing top quark jets from the light quark QCD background by
exploiting the jet substructure, without identifying the decay products. 

In our work~\cite{paper} we find that for a measurement at LHC of the top quark 
charge asymmetry it is not necessary to select events 
with very large invariant masses of the top-antitop quark pairs. 
We show that the highest statistical significance occurs with 
moderate selection cuts. Indeed, we find that the measurement 
of the charge asymmetry induced by QCD is better suited in the 
region of low top-antitop quark pair invariant masses. The 
higher statistics in this region compensates the smallness 
of the charge asymmetry. We also investigate the charge 
asymmetry generated by the exchange of a heavy color-octet 
resonance. We study the scenario where the massive extra 
gauge boson have arbitrary flavour independent vector and axial-vector 
couplings to quarks. This includes the case of the axigluon that 
we have already analyzed in Ref.~\cite{Antunano:2007da}.


\section{QCD induced charge asymmetry at LHC}
\label{sec:QCDLHC}

Top quark production at LHC is forward--backward symmetric in the 
laboratory frame as a consequence of the symmetric colliding 
proton-proton initial state. The charge asymmetry can be studied 
nevertheless by selecting appropriately chosen kinematic regions.
The production cross section of top quarks is, however,  
dominated by gluon-gluon fusion and thus the charge asymmetry 
generated from the $q\bar{q}$ and $gq$ ($g\bar{q}$) reactions is
small in most of the kinematic phase-space. 

\begin{figure}[th]
\begin{center}
\epsfig{file=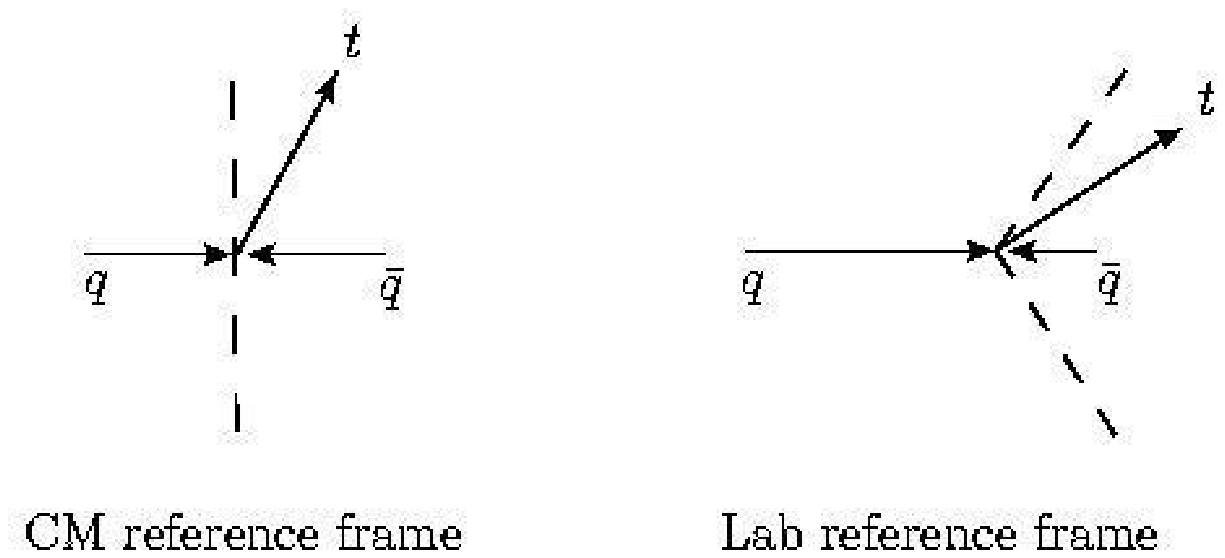,width=9cm} \\
\epsfig{file=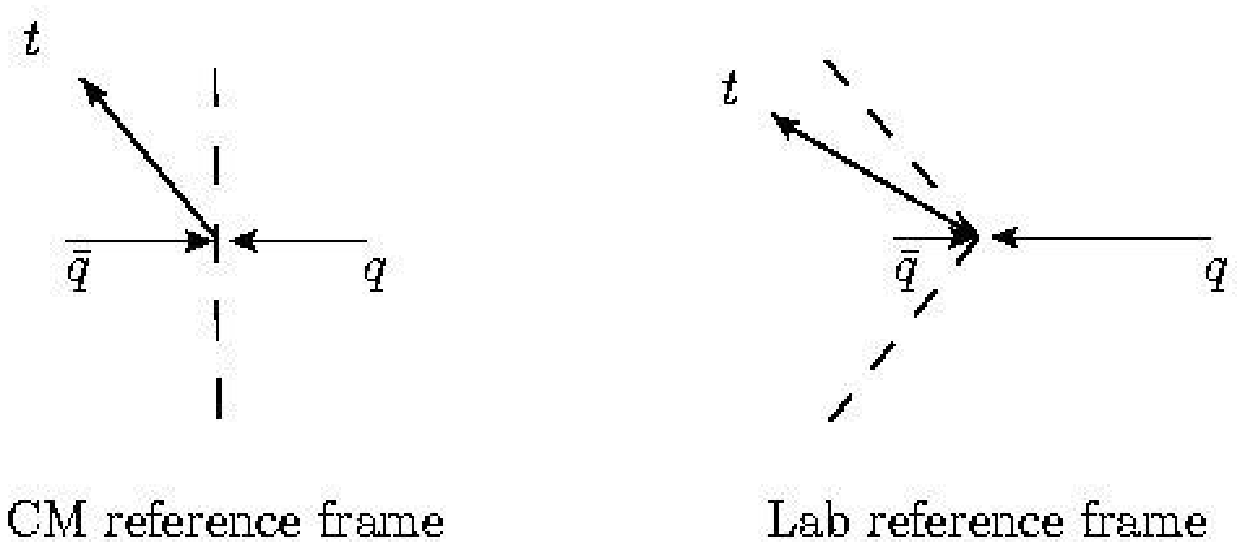,width=9cm}
\caption{Boost from the center-of-mass quark--antiquark reference frame to the laboratory frame.}
\label{fig:asym1}
\end{center}
\end{figure}

Nonetheless, QCD predicts at LHC a slight preference 
for centrally produced antitop quarks, with top quarks more abundant 
at very large positive and negative rapidities~\cite{mynlo}. 
The difference between the single particle inclusive distributions 
of $t$ and $\bar {t}$ quarks can be understood easily. 
Due to the proton composition in terms of quarks, production of $t\bar{t}(g)$ is dominated by initial quarks with 
large momentum fraction and antiquarks with small momentum 
fraction. QCD predicts that top (antitop) quarks are preferentially 
emitted in the direction of the incoming quarks (antiquarks)
in the partonic rest frame as shown in Fig.~\ref{fig:asym1}~(left graphs).
The boost into the laboratory frame ''squeezes'' the top mainly in the forward and backward directions, while antitops are left more abundant in the central region (see Fig.~\ref{fig:asym1}, right graphs).

We select events in a given range of rapidity $y$
and define the integrated charge asymmetry in 
the central region as~\cite{Antunano:2007da}:
\beq
A_C(y_C) = \frac{N_t(|y|\le y_C)-N_{\bar{t}}(|y|\le y_C)}
{N_t(|y|\le y_C)+N_{\bar{t}}(|y|\le y_C)}~.\label{eq:acyc}
\eeq
The central asymmetry $A_C(y_C)$ obviously vanishes if the 
whole rapidity spectrum is integrated, while a non-vanishing 
asymmetry can be obtained over a finite interval of rapidity. Since more antitop quarks are present in the central region (low $y$), we expect the asymmetry to be negative.
We also perform a cut on the invariant mass of the top-antitop quark 
pair, $m_{t\bar t} > m_{t\bar t}^{\rm{min}}$,
because that region of the phase space is more sensitive 
to the quark-antiquark induced events rather than the 
gluon-gluon ones, so that the asymmetry is enhanced.
The main virtue of the central asymmetry is that it vanishes 
exactly for parity-conserving processes.

\begin{figure}[th]
\begin{center}
\epsfig{file=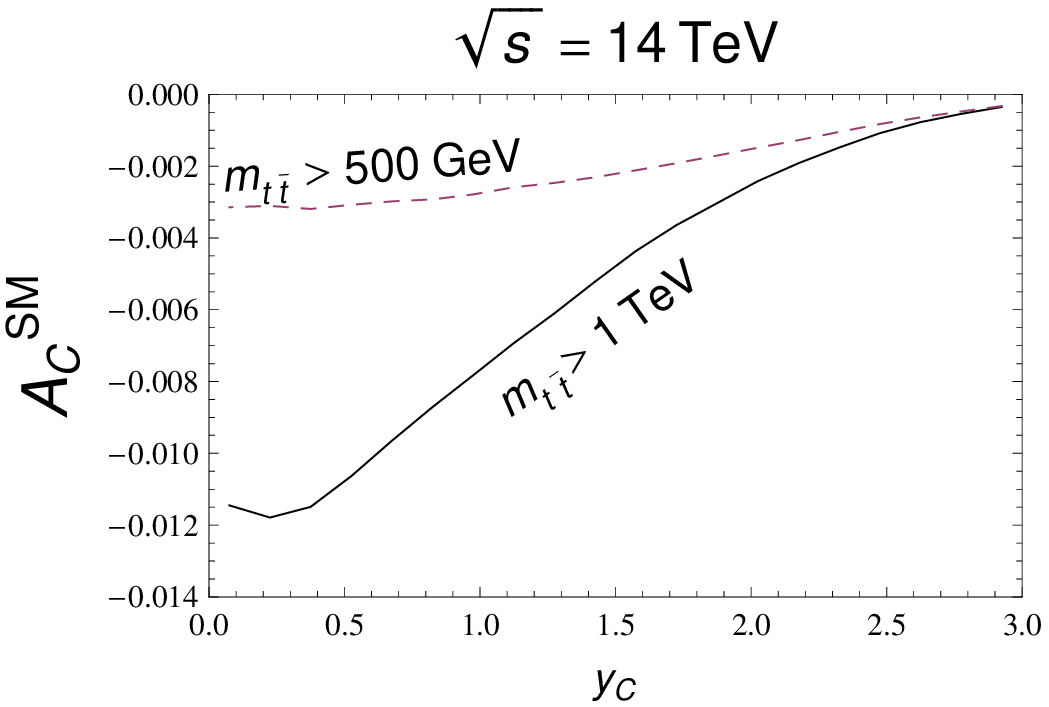,width=7cm,height=5cm} 
\epsfig{file=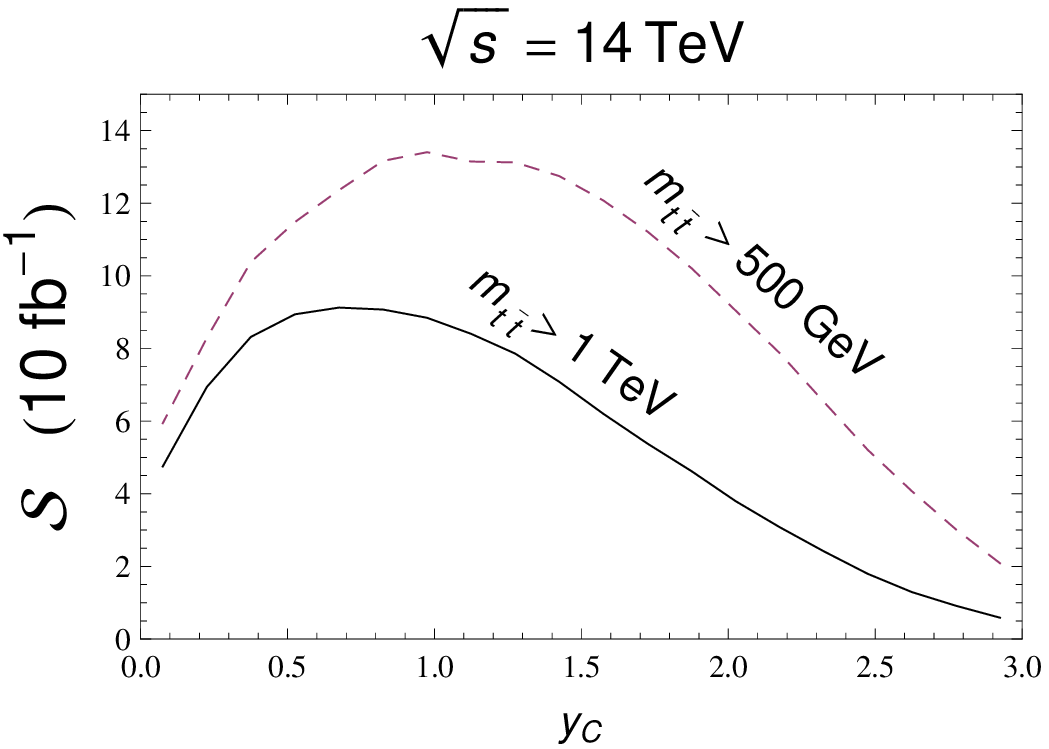,width=7cm,height=5cm}  
\caption{Central charge asymmetry at LHC as predicted by QCD, as a function of 
the maximum rapidity $y_C$ (left plot), and corresponding statistical 
significance (right plot), for two different cuts on the top-antitop 
quark pair invariant mass.}
\label{fig:LHC_QCD_yc}
\end{center}
\end{figure}

In Fig.~\ref{fig:LHC_QCD_yc}~(left plot) we show the central charge asymmetry 
at $14$~TeV as a function of the maximum rapidity $y_C$ 
for two different values 
of the cut on the invariant mass of the top-antitop quark pair
$m_{t\bar t}> 500~$GeV, and $1$~TeV, respectively. 
As expected, the central charge asymmetry is negative, is 
larger for larger values of the cut $m_{t\bar t}^{\rm{min}}$, and vanishes 
for large values of $y_C$. 
We also show in Fig.~\ref{fig:LHC_QCD_yc}~(right plot) 
the corresponding statistical significance ${\cal S}$ 
of the measurement, defined as 
\beq
{\cal S}^{\rm{SM}} = A_C^{\rm{SM}} \, \sqrt{(\sigma_t+\sigma_{\bar t})^{\rm{SM}} 
\, {\cal L}} = \frac{N_t-N_{\bar t}}{\sqrt{N_t+N_{\bar t}}}~, 
\eeq
where ${\cal L}$ denotes the total integrated luminosity for which we 
take ${\cal L}=10$~fb$^{-1}$ at $\sqrt{s}=14$~TeV, accordingly to the first LHC schedule,. 
The maximum significance is reached at $y_C=1$ for $m_{t\bar t}> 500~$GeV, and at 
$y_C=0.7$ for $m_{t\bar t}>1$~TeV. Surprisingly, although the 
size of the asymmetry is greater for the larger value of 
$m_{t\bar t}^{\rm{min}}$, its statistical significance is higher 
for the lower cut. This is a very interesting 
feature because softer top and antitop quarks should be identified 
more easily than the very highly boosted ones.

\begin{figure}[th]
\begin{center}
\epsfig{file=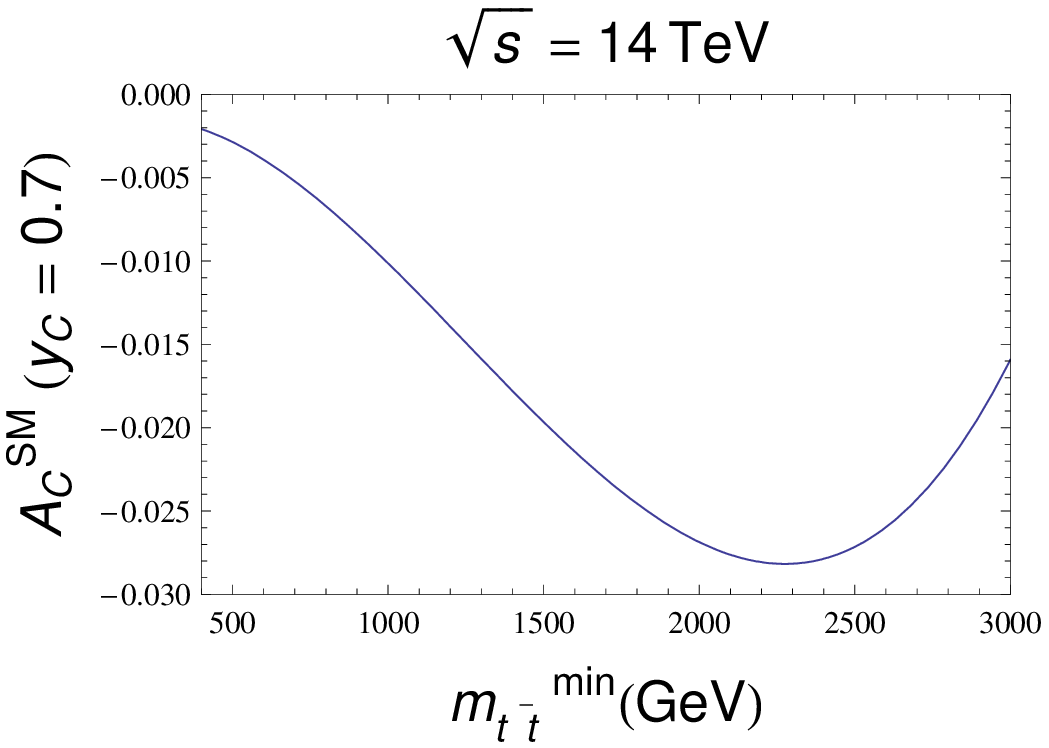,width=7cm,height=5cm} 
\epsfig{file=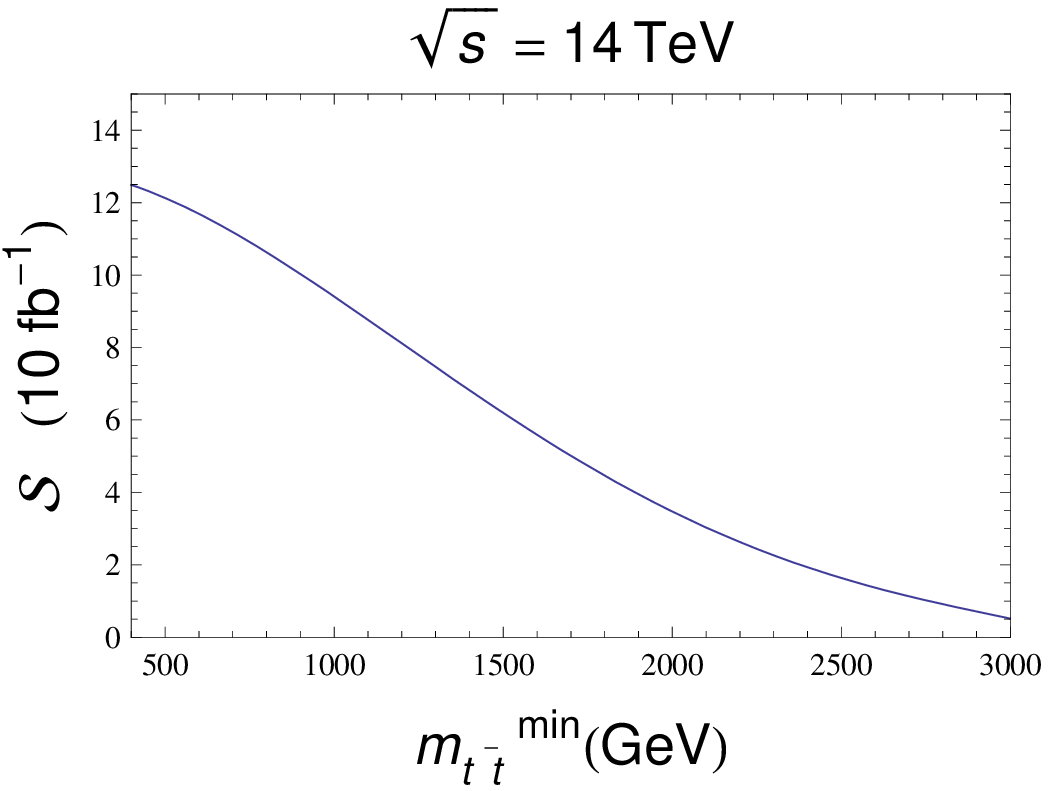,width=7cm,height=5cm}  
\caption{Central charge asymmetry and statistical significance 
at LHC from QCD, as a function of the cut $m_{t\bar t}^{\rm{min.}}$, 
for $y_C=0.7$.}
\label{fig:LHC_QCD_mtt_14TeV}
\end{center}
\end{figure}

We now fix the value of the maximum rapidity to $y_C=0.7$ and study 
the size of the asymmetry and its statistical significance as a 
function of $m_{t\bar t}^{\rm{min}}$.
Our results are shown in Fig.~\ref{fig:LHC_QCD_mtt_14TeV}
for $\sqrt{s}=14$~TeV and ${\cal L}=10$~fb$^{-1}$. We find that the asymmetry 
increases for larger values of $m_{t\bar t}^{\rm{min}}$, 
while the statistical significance is larger without 
introducing any selection cut. Note that the size of the asymmetry 
decreases again above $m_{t\bar t}^{\rm{min}}=2.5$~TeV
because in that region the $gq(\bar q)$ events compensate 
the asymmetry generated by the $q\bar q$ events; their 
contributions are of opposite sign. 
Although we have not taken into account experimental 
efficiencies, we can conclude that $10$~fb$^{-1}$ of data at the design 
energy of the LHC seems to be enough for a clear measurement of the 
QCD asymmetry.

\section{Charge asymmetry of color-octet resonances at LHC}
\label{sec:octetLHC}

We study here the charge asymmetry produced at LHC 
by the decay of a color-octet resonance $G$ to top quarks, 
in the scenario where the vector $g_V^{q}$ and 
axial-vector $g_A^{q}$ couplings are flavour independent. 
We evaluate the central asymmetry in \Eq{eq:acyc}, and 
its statistical significance, defined as~\cite{Djouadi:2007eg}
\beq
{\cal S}^{\rm G} = 
\frac{A_C^{\rm{G+SM}}-A_C^{\rm{SM}}}{\sqrt{1-(A_C^{\rm{SM}})^2}} 
\, \sqrt{(\sigma_t+\sigma_{\bar t})^{\rm{SM}} \, {\cal L}}~,
\eeq
for different values of the couplings and the kinematical cuts.


The corresponding differential cross section is given 
in \Eq{eq:bornqq} of the Appendix.
The charge asymmetry is built up from the two contributions
of the differential partonic cross section that are odd in the polar angle.
The first one arises from the interference with the gluon amplitude, 
and is proportional to the product of the axial-vector couplings of the 
light and the top quarks. This contribution, provided that the product 
of couplings is positive, is negative in the forward direction 
for invariant masses of the top-antitop quark pair below the resonance mass, 
and changes sign above. At LHC, because of the boost into the laboratory frame (cf. discussion in Section~\ref{sec:QCDLHC}), this means that a positive central asymmetry is found for values of the cut in 
the invariant mass of the top-antitop quark pair below the mass of the 
resonance and a negative asymmetry above. This means that the asymmetry has to vanish at a certain intermediate 
value of that cut, close to and below the resonance mass. 
  
The second contribution, arising from the squared amplitude of the heavy 
resonance, although always positive for positive couplings,
is suppressed with respect to the contribution of the interference 
term by two powers of the resonance mass.
For large values of the vector couplings,
however, it might compensate the interference contribution, 
then leading to a positive contribution in the forward region.

In summary, we expect to find two maxima in the statistical 
significance as a function of $m_{t\bar{t}}^{\rm{min}}/m_{G}$.
Starting from the threshold, where the asymmetry is small because
gluon-gluon fusion dominates there, the size of the 
central asymmetry will grow by increasing $m_{t \bar t}^{\rm{min}}$,
as the quark-antiquark annihilation process becomes more
and more important. Since the asymmetry induced by the excited gluon 
will vanish at a certain critical point, its statistical significance
will do as well, and will reach a maximum at an intermediate value 
between that critical point and the threshold. 
Above the critical point, the asymmetry becomes negative and 
its statistical significance increases again, until the event yield 
becomes too small. A second maximum in the statistical significance 
will be generated there. For certain values of the vector couplings, however, the contribution from the squared amplitude 
of the exotic resonance is greater than the interference term. In this case, the central asymmetry generated by 
the exotic resonance will be negative exclusively, 
and we will find only one maximum in the 
statistical significance.

\begin{figure}[th]
\begin{center}
\epsfig{file=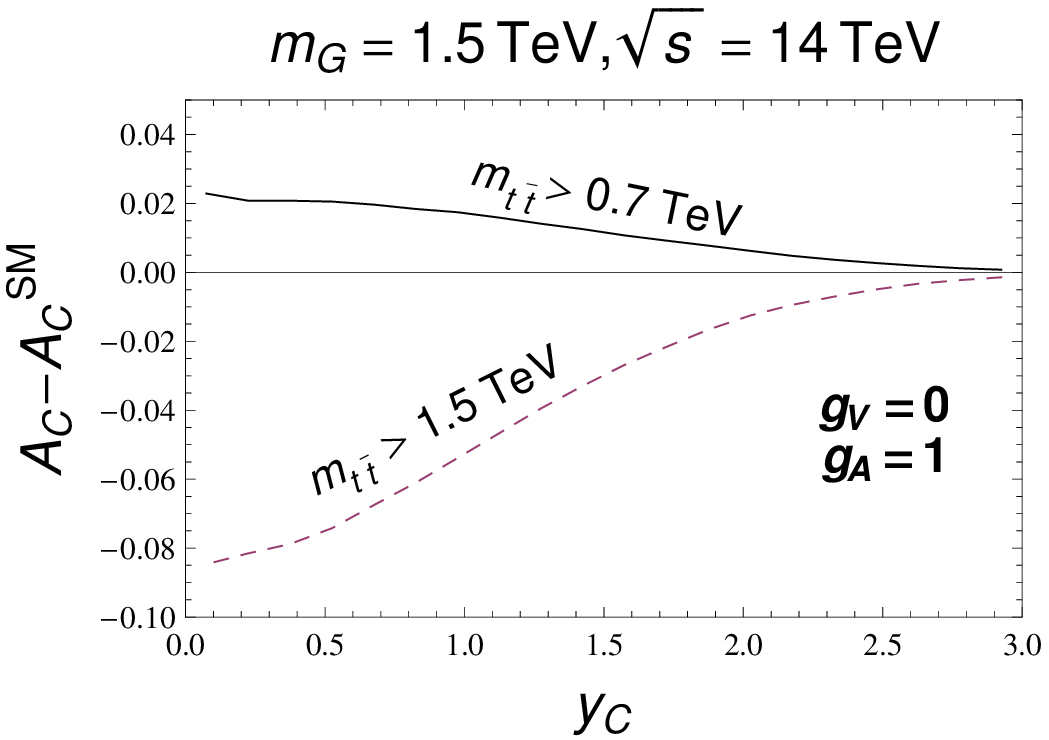,width=7cm,height=5cm}
\epsfig{file=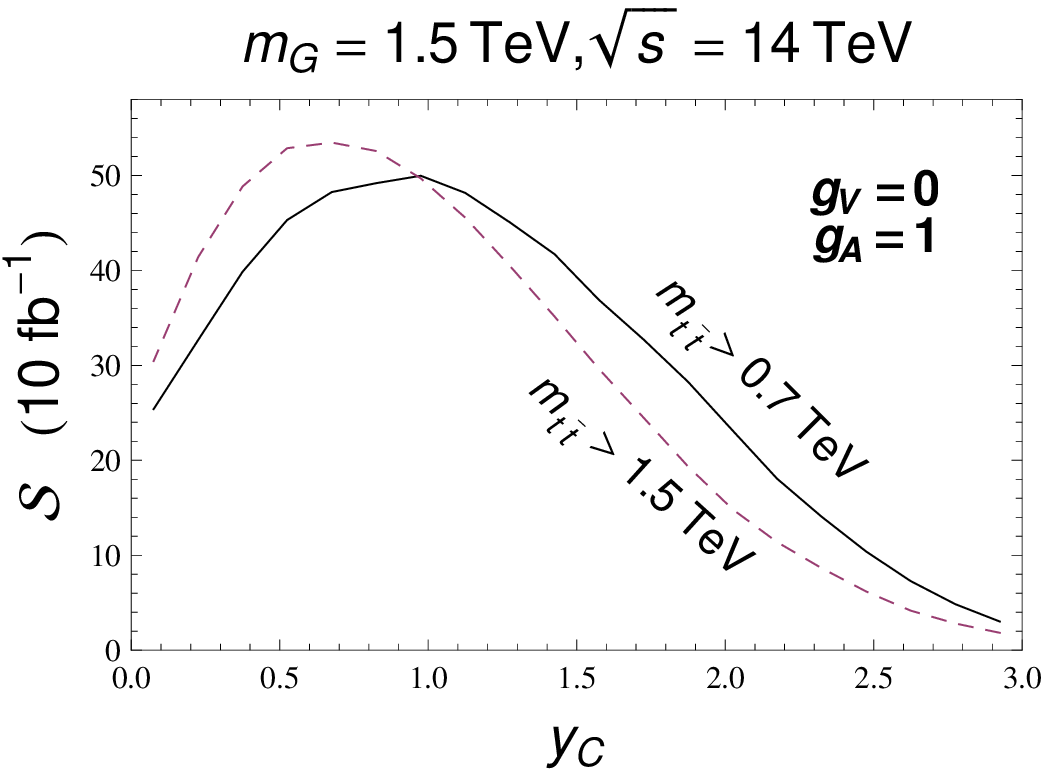,width=7cm,height=5cm}
\caption{Central charge asymmetry (left plot) and statistical 
significance (right plot) at LHC as a function of the maximum 
rapidity, for two different cuts on the 
top-antitop quark invariant mass. The axigluon case is shown.}
\label{fig:LHC_KK_yc_14TeV}
\end{center}
\end{figure}

In our first analysis we shall determine the value of 
the maximum rapidity $y_C$ that maximizes the statistical significance. 
We fix the resonance mass at $1.5$ TeV,
and impose two different cuts on the invariant mass 
of the top-antitop quark pair, namely $m_{t\bar t}> 700 \;\mathrm{GeV}$ 
and $m_{t\bar t}> 1.5\; \mathrm{TeV}$. 
We choose as an example the axigluon case: $g_V^{q(t)}=0$, $g_A^{q(t)}=1$.
In Fig \ref{fig:LHC_KK_yc_14TeV}, 
we present the results obtained for the central asymmetry and the 
statistical significance 
at $14$ TeV centre-of-mass energy. 
We notice that the central asymmetry suffers a change 
of sign by passing from the lower cut to the higher one. 
This means that it will vanish for a given value of the cut, 
thus making the statistical significance vanishing also. 

\begin{figure}[th]
\begin{center}
\epsfig{file=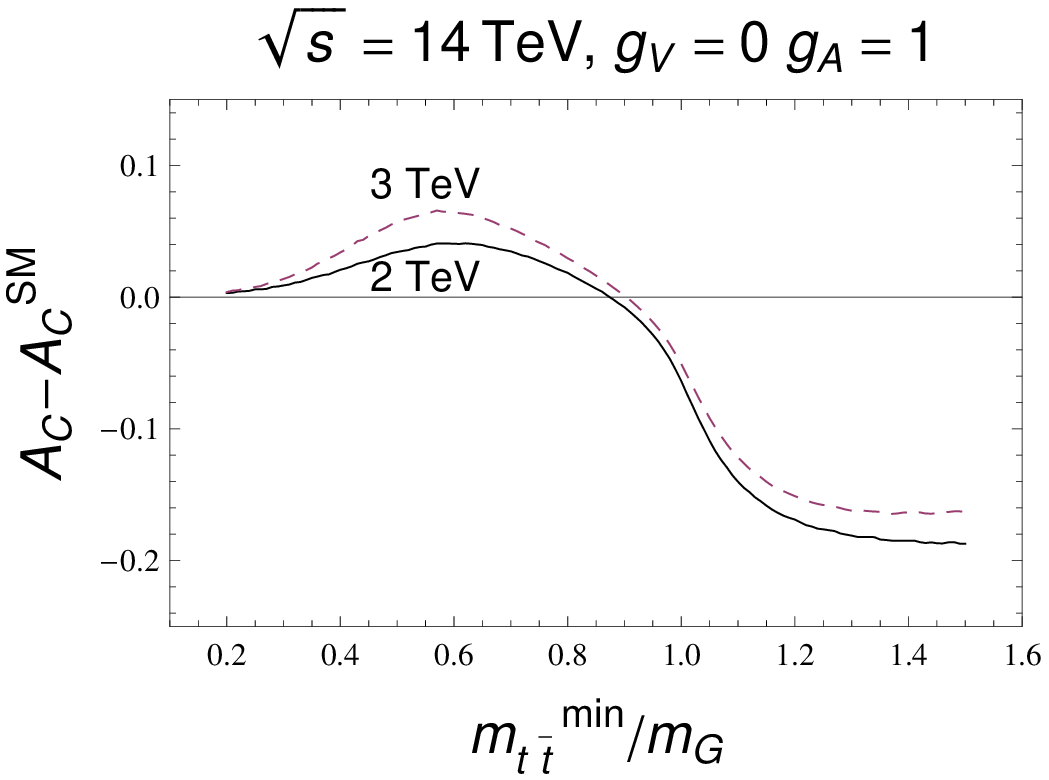,width=6.2cm} 
\epsfig{file=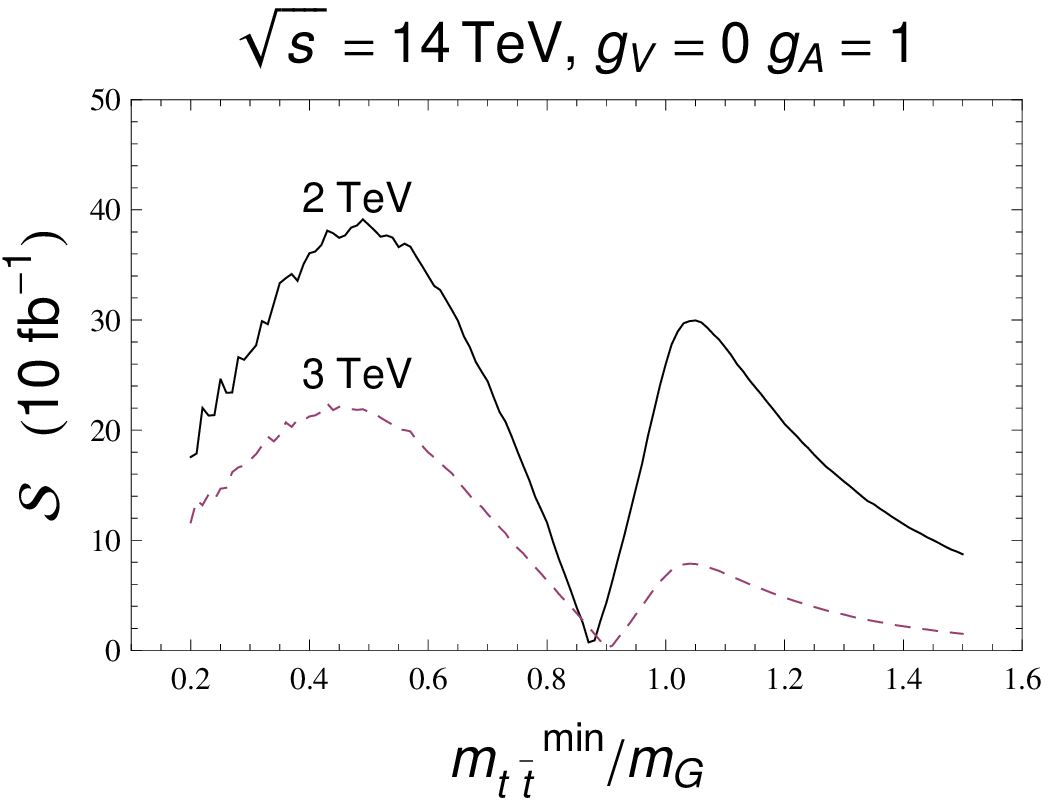,width=6cm} 
\caption{Central asymmetry and statistical significance at LHC, for $g_V=0$, $g_A=1$, as a function of the cut on the top-antitop quark pair invariant mass for $m_G=2$ and $3$~TeV.}
\label{fig:LHC_KK_gvga_14TeV}
\end{center}
\end{figure}

By looking at the corresponding significance we find that 
$y_C=0.7$ is a good choice for any value of the couplings.
Thus, we use this value to find the best cut for the
top-antitop quark pair invariant mass. 
In order to do that, we choose several 
values of the parameters and we study the trend of the 
significance as a function of $m_{t\bar t}^{\rm{min}}/m_{G}$. 
We show again the results for the axigluon case, in Fig \ref{fig:LHC_KK_gvga_14TeV}. The optimal cuts depend, 
of course, on the values of the vector and axial-vector 
couplings, but either $m_{t\bar{t}}^{\rm{min}}/m_{G}=0.5$ 
or $m_{t\bar{t}}^{\rm{min}}/m_{G}=0.8$ provide a reasonable 
statistical significance for almost all the 
combinations of the couplings. 
This is an important result, because it means that a relatively 
low cut -- at about half of the mass of the resonance or even below -- 
is enough to have a good statistical significance, 
and a clear signal from the measurement of the charge asymmetry.

\begin{figure}[th]
\begin{center}
\epsfig{file=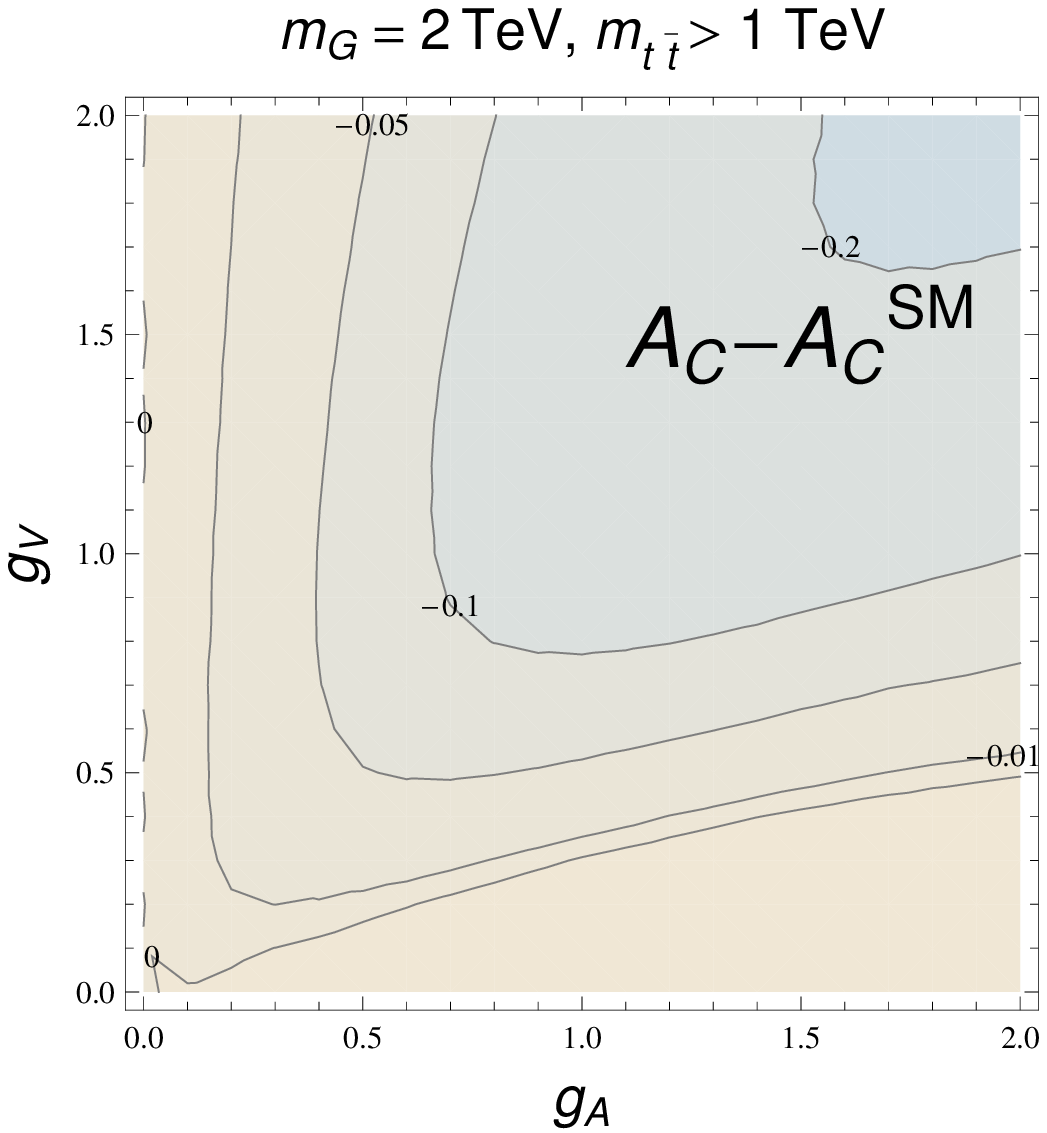,width=5cm} 
\epsfig{file=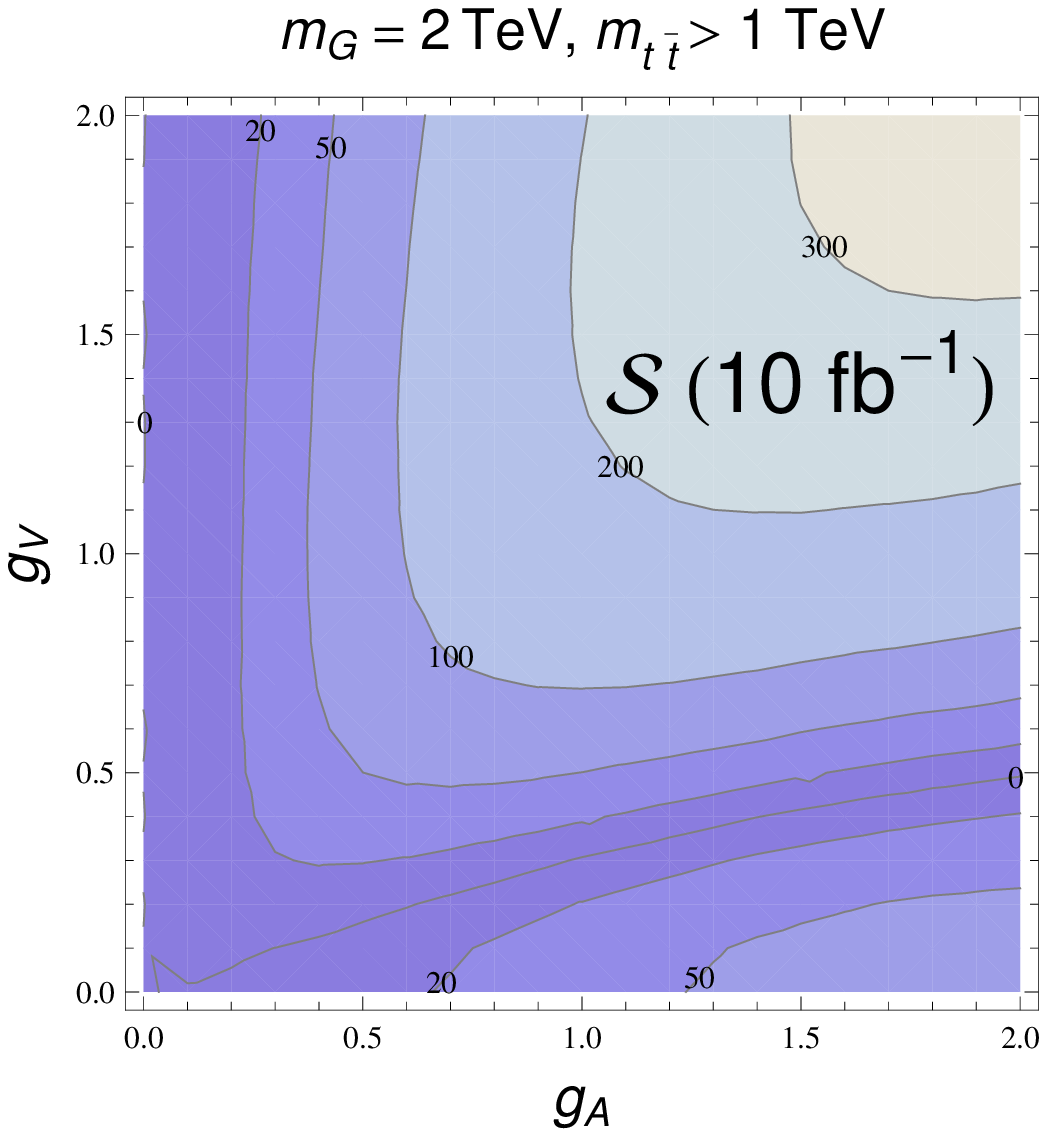,width=5cm} \\
\epsfig{file=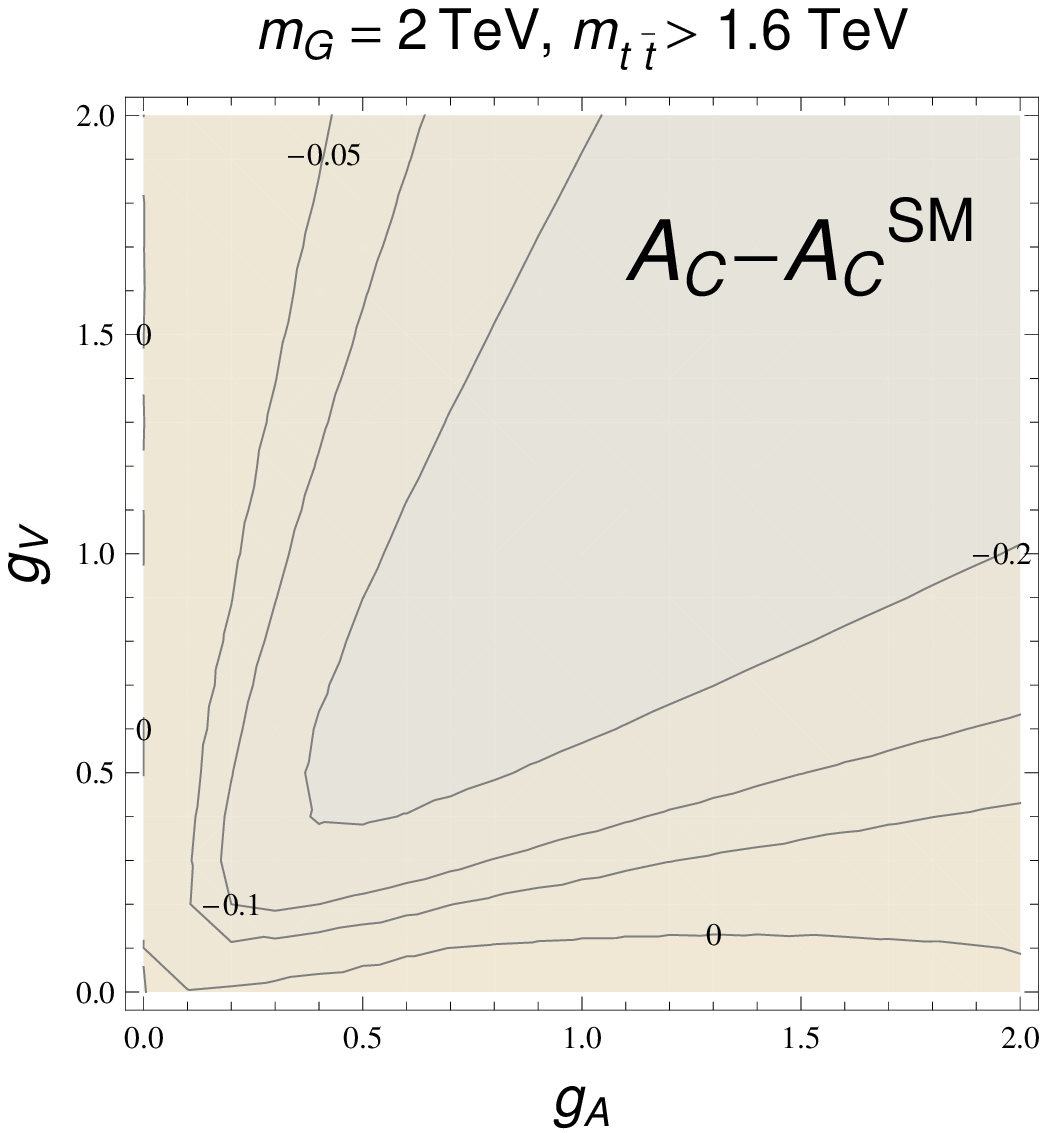,width=5cm}
\epsfig{file=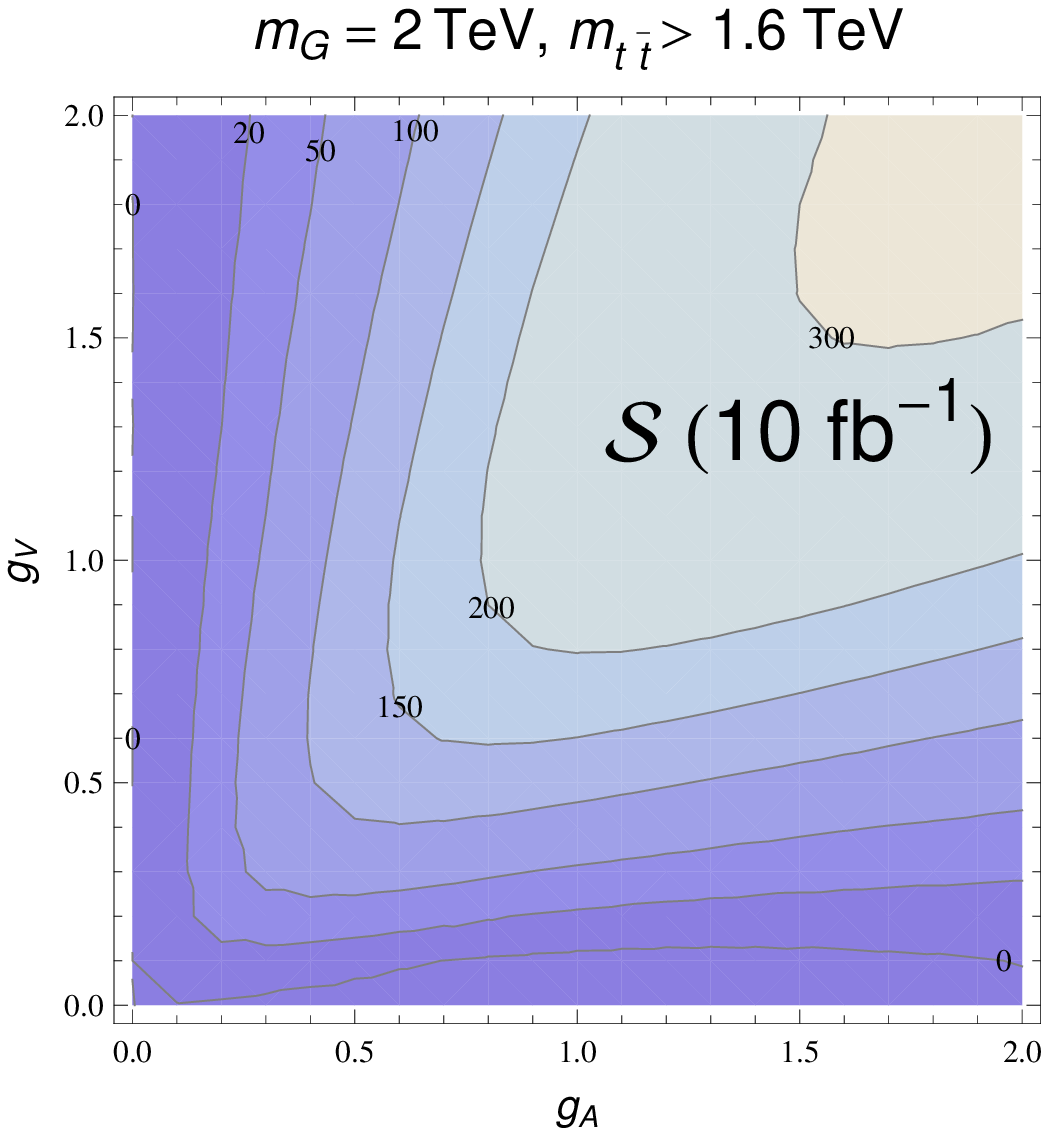,width=5cm} \\
\epsfig{file=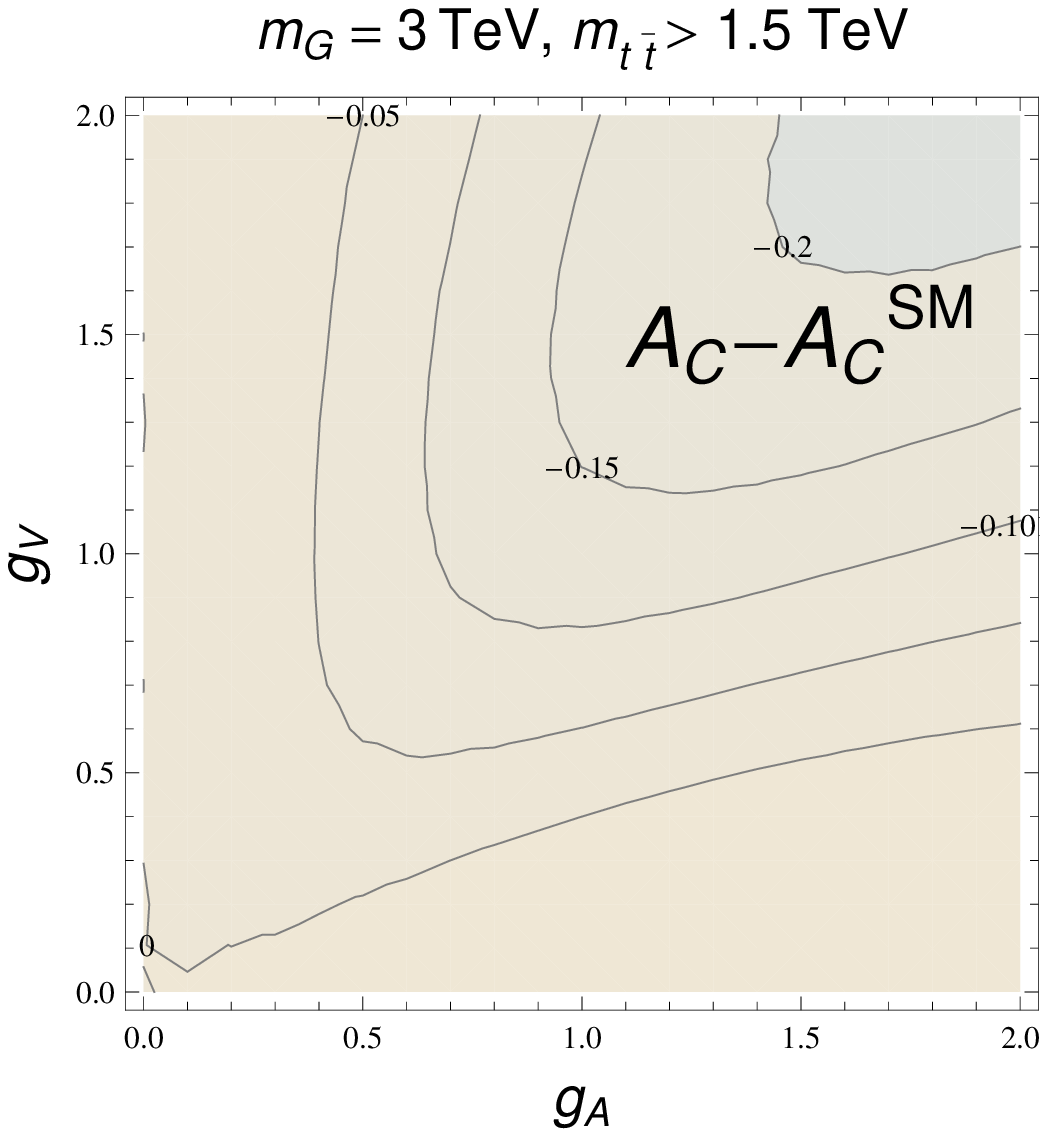,width=5cm} 
\epsfig{file=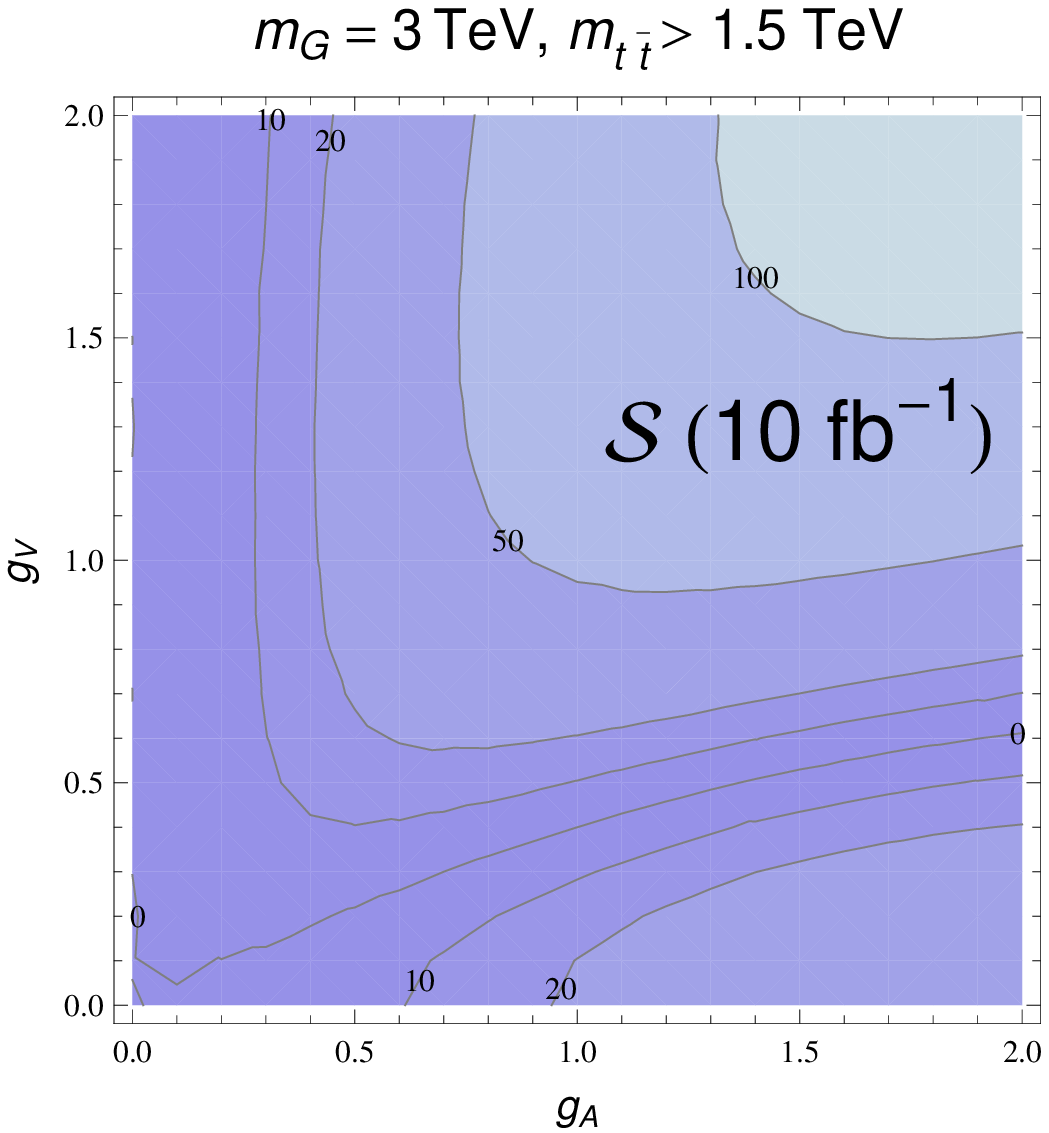,width=5cm}\\
\epsfig{file=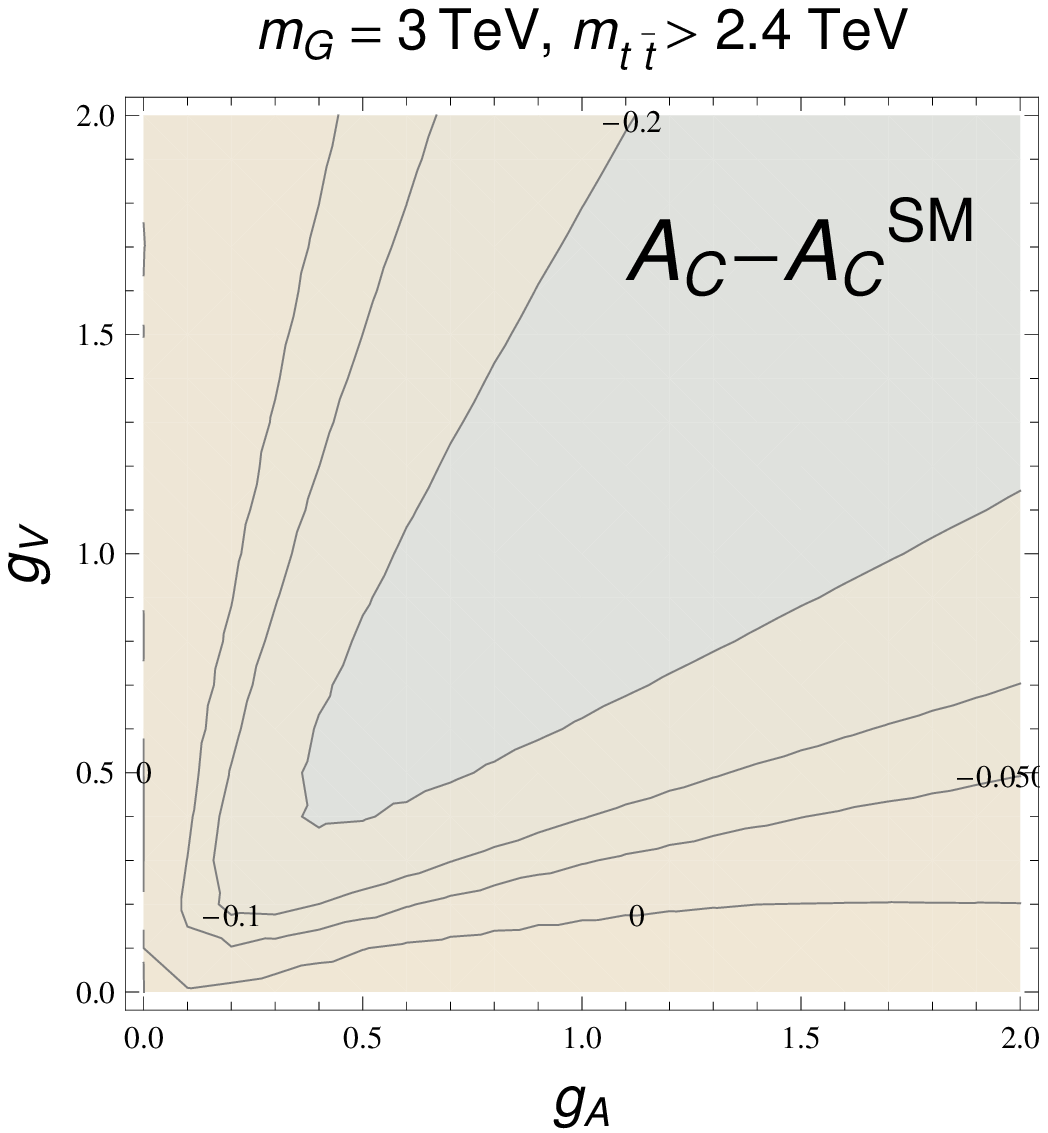,width=5cm} 
\epsfig{file=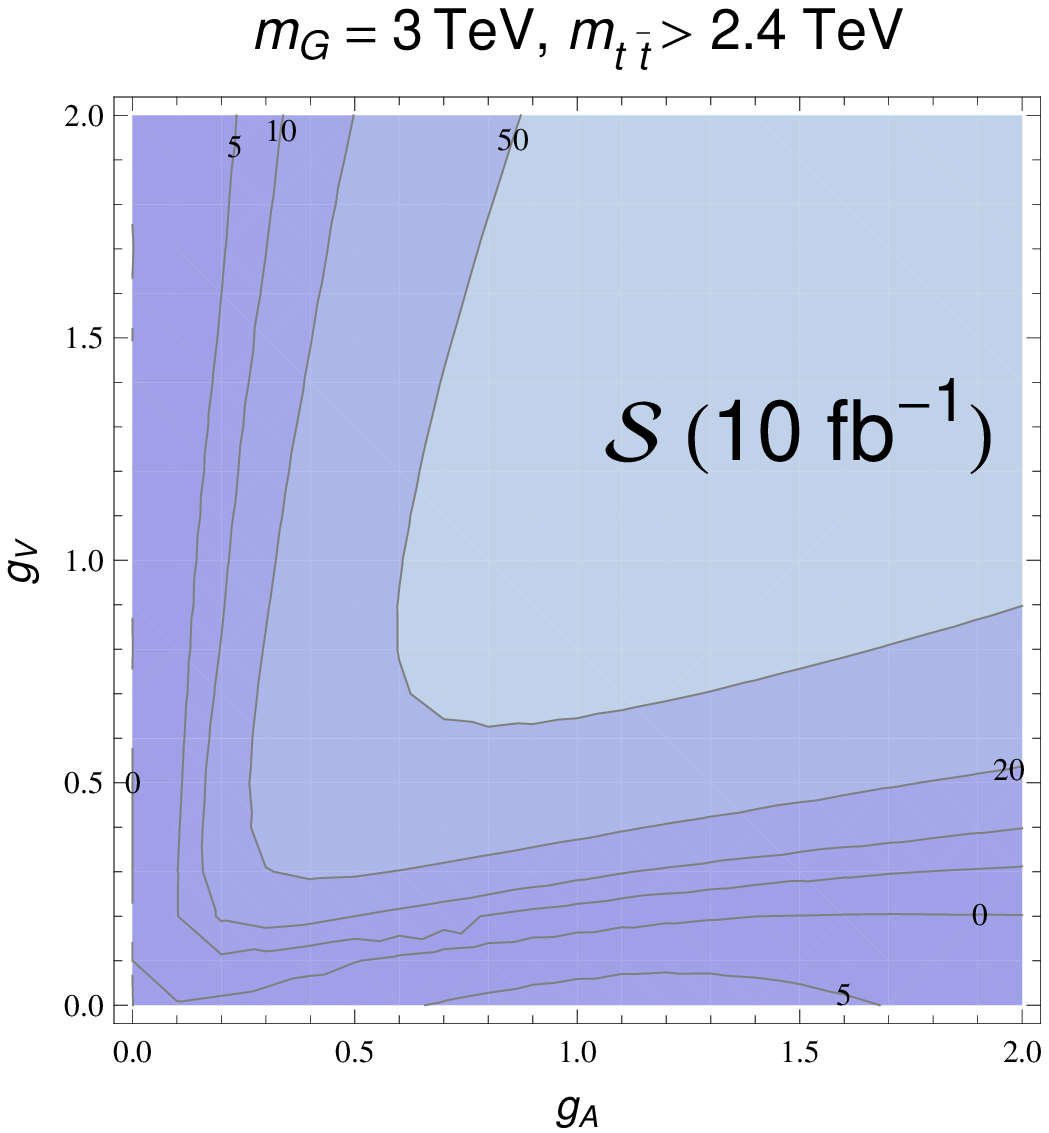,width=5cm} 
\caption{Central charge asymmetry and statistical significance 
at LHC for $14$ TeV energy in the $g_A$-$g_V$ plane, 
for different values of the resonance mass and the cut
on the top-antitop quark pair invariant mass.}
\label{fig:LHC_KK_2d_14TeV}
\end{center}
\end{figure}

We now fix $m_{t\bar{t}}^{\rm{min}}/m_{G}=0.5$ and 
$m_{t\bar{t}}^{\rm{min}}/m_{G}=0.8$, and we study how the central
asymmetry and its statistical significance vary as a function of
the vector and the axial-vector couplings, 
for a given value of the resonance mass.  
These choices, for which we have found the best 
statistical significances, are of course arbitrary and are 
not necessarily the best for all the values of the vector
and axial-vector couplings. For illustrative purposes are, 
however, good representatives. 
We have chosen $m_{G}=1.5, 2$ and $3$~TeV. 
The results in the $(g_V,g_A)$ plane are presented in Fig
\ref{fig:LHC_KK_2d_14TeV}. 
It is possible to see that the pattern of the size of the asymmetry 
is quite similar independently of the value of the resonance mass;
it depends mostly on the ratio $m_{t\bar{t}}^{\rm{min}}/m_{G}$.
A sizable asymmetry is found whatever 
the value of the resonance mass is. The statistical significance, 
as expected, decreases with higher resonance masses.

\section{Conclusions}

We have analyzed at LHC the charge asymmetry of top-antitop quark pair production in QCD and
through the exchange of a color-octet heavy boson with flavor independent
coupling to quarks. 

This is an
observable that is very sensitive to new physics. 
We have studied the statistical significance of the 
measurement of such an asymmetry, and we have found that  
it is possible to tune the selection cuts in order to find a 
sensible significance.
The maximum of the statistical significance for the measurement 
of the asymmetry as predicted by QCD is obtained without introducing 
any cut in the invariant mass of the top-antitop quark pair, 
even if the asymmetry is smaller in this case.

When a heavy resonance is considered, one or two maxima in the 
significance spectrum are found, depending on the size of the couplings. 
The position of the peaks depends on the ratio 
$m_{t\bar{t}}^{\mathrm{min}}/m_{G}$ and not on the resonance mass. 
One of the peaks can be located at an energy scale as low as one 
half of the resonance mass, or even below. 
Data samples of top and antitop quarks that are not too energetic 
can then be used to detect or exclude the existence of this 
kind of resonances. 

%
%

\ack

The work of P.F. is supported by an I3P Fellowship from 
Consejo Superior de Investigaciones Cient\'{\i}ficas (CSIC). This work is also partially supported by Ministerio de Ciencia e Innovaci\'on 
under Grants  No. FPA2007-60323 and CPAN (CSD2007-00042), Generalitat Valenciana under grant No. PROMETEO/2008/2009 and 
European Commission MRTN FLAVIAnet under Contract No.  
MRTN-CT-2006-035482.



\appendix

\section*{Appendix}
\label{ap:born}

\setcounter{section}{1}

The Born cross-section for $q\bar{q}$ annihilation in the presence of 
a color-octet vector resonance reads
\bea 
\frac{d\sigma^{q\bar{q}\rightarrow t \bar{t}}}{d\cos \hat{\theta}} &=& 
\alpha_s^2 \: \frac{T_F C_F}{N_C} \:  
\frac{\pi \beta}{2 \hat{s}}
\Bigg( 1+c^2+4m^2 + \frac{2 \hat{s} (\hat{s}-m_G^2)}
{(\hat{s}-m_G^2)^2+m_G^2 \Gamma_G^2} 
\left[ g_V^q \, g_V^t \, (1+c^2+4m^2) + 2 \, g_A^q \, g_A^t \, c  \right] 
\nn \\ &+&
\frac{\hat{s}^2} {(\hat{s}-m_G^2)^2+m_G^2 \Gamma_G^2} 
\bigg[ \left( (g_V^q)^2+(g_A^q)^2 \right)
\bigg( (g_V^t)^2 (1+c^2+4m^2) \nn \\ &+&  (g_A^t)^2 (1+c^2-4m^2) \bigg) 
+ 8 \, g_V^q \, g_A^q \, g_V^t \, g_A^t \, c \, \bigg]
\Bigg)~,
\label{eq:bornqq}
\eea
where $\hat{\theta}$ is the polar angle of the top quark with respect 
to the incoming quark in the center of mass rest frame, 
$\hat{s}$ is the squared partonic invariant mass,
$T_F=1/2$, $N_C=3$ and $C_F=4/3$ are the color factors,
$\beta = \sqrt{1-4m^2}$ is the velocity of the top quark, 
with $m=m_t/\sqrt{\hat{s}}$, and $c = \beta \cos \hat{\theta}$.
The parameters $g_V^q (g_V^t)$, $g_A^q(g_A^t)$ represent the vector
and vector-axial couplings among the excited gluons and the light quarks 
(top quarks). 

There are two terms in \Eq{eq:bornqq} that are odd in the polar 
angle and therefore there are two contributions to the charge asymmetry.
The first one arises from the interference of the SM amplitude with 
the resonance amplitude, and the second one 
from the square of the resonance amplitude. 
The former depends on the axial-vector couplings only, while 
the latter is proportional to both the vector and the axial-vector
couplings. For large values of the resonance mass, the second term is 
suppressed, and the charge asymmetry will depend mostly on the 
value of the axial-vector couplings, and residually on the vector 
couplings. The decay width is given by:
\bea
\Gamma_G &\equiv& \sum_q \Gamma (G \to q\overline{q}) 
\approx \frac{\alpha_{s}\, m_G \, T_F}{3} 
\Bigg[\sum_q \left( (g_V^q)^2+(g_A^q)^2 \right) \nn \\
&& +\sqrt{1-\frac{4m_t^{2}}{m_G^{2}}}
\left( (g_V^t)^2 \left(1+\frac{2m_t^{2}}{m_G^{2}}\right) 
+ (g_A^t)^2 \left(1-\frac{4m_t^{2}}{m_G^{2}}\right) 
\right) \Bigg]~.
\eea
We assume that the Born gluon-gluon fusion cross-section 
is the same as in the SM:
\beq
\frac{d\sigma^{gg\rightarrow t \bar{t}}}{d\cos \hat{\theta}} = 
\alpha_s^2 \: \frac{\pi \beta}{2 \hat{s}}  
\left(\frac{1}{N_C(1-c^2)}-\frac{T_F}{2C_F}\right)
\left(1 + c^2 +8 m^2-\frac{32 m^4}{1-c^2}\right)~.
\eeq


\end{document}